\documentclass{amsproc}

\newtheorem{theorem}{Theorem}[section]

\newtheorem{proposition}[theorem]{Proposition}
\theoremstyle{definition}

\newtheorem{example}[theorem]{Example}
\theoremstyle{remark}

\newtheorem{remarks}[theorem]{Remarks}
\numberwithin{equation}{section}
\newcommand{\restr}{\vert\hskip -5.5pt \phantom{\vert}^{\scriptscriptstyle
            \backslash}}
\newcommand{\QED}{\mbox{\rule[-1.0pt]{5pt}{8pt}}}

\begin{document}

\title{Point interactions in a tube}

\author{Pavel Exner}
\address{Nuclear Physics Institute, Academy of Sciences,
25068 \v Re\v z near Prague; Doppler Institute, Czech Technical
University, B\v rehov{\'a} 7, 11519 Prague, Czech Republic}
\curraddr{} \email{exner@ujf.cas.cz}
\thanks{The research has been partially supported by GAAS under the
contract A1048801.}



\begin{abstract}
We discuss discuss spectral and scattering properties of a
particle confined to a straight Dirichlet tube in $\mathbb{R}^3$
with a family of point interactions.
\end{abstract}
\maketitle

\noindent Point interactions belong to the list of problems to
which Sergio Albeverio made a significant contribution. This topic
combines a practical importance as a source of numerous solvable
models with an aesthetic appeal as the monograph \cite{AGHH}
witnesses. At the same time it is far to be closed; despite the
extensive and thorough character of the mentioned treatise new
questions still arise.

One of them concerns point interactions in tubular regions which
represent a natural model for a ``quantum wire" with impurities.
The simplest situation when the tube is a straight planar strip
was investigated in \cite{EGST}; we refer to this paper for a
detailed motivation and bibliography. In the present paper we
present a brief discussion of a straight tube in $\mathbb{R}^3$
with a family of point interactions.

\section{The one-center case}

\noindent Let $\Omega:= \mathbb{R}\times M$ where
$M\subset\mathbb{R}^2$ is a closed compact set; we suppose that it
is pathwise connected and $\partial M$ has the segment property
\cite{RS}. The free Hamiltonian is the corresponding Dirichlet
Laplacian, $H_0= -\Delta_D^{\Omega}$ with the domain
$W^{2,2}_0(\Omega)$. It can be expressed by means of the
one-dimensional Laplacian and $-\Delta_D^M$. The last named
operator has a purely discrete spectrum; we denote by $\chi_n,\:
\nu_n$ its eigenfuctions and eigenvalues, respectively. For any
$z\in\mathbb{C}\setminus [\nu_0,\infty)$ the free resolvent is an
integral operator with the kernel
\begin{equation} \label{free kernel}
G_0(\vec x_1,\vec x_2;z)\,\equiv\, (H_0\!-z)^{-1}(\vec x_1,\vec
x_2)\,=\, {i\over 2}\, \sum_{n=0}^{\infty}\,
\frac{e^{ik_n(z)|x_1-x_2|}}{k_n(z)}\, \chi_n(\vec y_1) \chi_n(\vec
y_2)\,,
\end{equation}
where $\vec x_j=(x_j,\vec y_j)$ and $k_n(z):=\,
\sqrt{z\!-\!\nu_n}$, which is defined and smooth except at $\vec
x_1= \vec x_2$. It is a multivalued function of $z$ with cuts
$[\nu_n,\infty), \: n=0,1,\dots$.

Suppose now that a point interaction is situated at $\vec a=
(a,\vec b) \in M^o$. We define it as in \cite{AGHH}, i.e. as a
self-adjoint extension of the operator $-\Delta_D^\Omega \restr
C_0^{\infty}(\Omega\setminus\{\vec a\})$. We employ generalized
boundary values
\begin{equation} \label{generalized bv}
L_0(\psi,\vec a)\,:=\, \lim_{\vec x\to\vec a}\, \psi(\vec x) |\vec
x\!-\!\vec a|\,, \quad\; L_1(\psi,\vec a)\,:=\, \lim_{\vec
x\to\vec a} \left\lbrack \psi(\vec x)- {L_0(\psi,\vec a)\over
|\vec x\!-\!\vec a|} \right\rbrack\,;
\end{equation}
then the extension in question is specified by the boundary
condition
\begin{equation} \label{bc}
L_1(\psi,\vec a)+4\pi\alpha L_0(\psi,\vec a)\,=\,0
\end{equation}
for a given $\alpha\in\mathbb{R}$. We shall denote it
$H(\alpha,\vec a)$; the case $\alpha=\infty$, i.e. $L_0(\psi,\vec
a)=0$, corresponds to the free Hamiltonian $H_0$.

The resolvent of $H(\alpha,\vec a)$ is obtained by Krein's
formula. Mimicking the argument of \cite {EGST} we get
\begin{equation} \label{Krein}
(H(\alpha,\vec a)\!-\!z)^{-1}(\vec x_1,\vec x_2)\,=\, G_0(\vec
x_1,\vec x_2;z) +\, {G_0(\vec x_1,\vec a;z) G_0(\vec a,\vec
x_2;z)\over \alpha-\xi(\vec a;z)}\,,
\end{equation}
where $\xi(\vec a;z)$ is the regularized Green's function at $\vec
a$,
\begin{equation} \label{xi}
\xi(\vec a;z)\,=\, \lim_{u\to 0}\, \left\lbrack {i\over 2}\,
\sum_{n=0}^{\infty}\, {e^{ik_n u}\over k_n} |\chi_n(\vec b)|^2 -\,
{1\over 4\pi u}\, \right\rbrack\,.
\end{equation}
The existence of the limit follows from the kernel behaviour at
the singularity \cite{Ti}. However, we also need a prescription
how to compute it and this differs from the two-dimensional case.
We use semiclassical properties of the above series terms
\cite{RS}. The transverse eigenvalues behave as $\nu_n\approx
4\pi|M|^{-1}n$ for $n\to\infty$ so $k_n= 2i\sqrt{\pi}|M|^{-1/2}
\sqrt{n}+ \mathcal{O}(1)$. On the other hand, the probability
densities $|\chi_n|^2$ are rapidly oscillating functions. Since
$M$ supports no potential the mean value of these oscillations
equals the constant $|M|^{-1}$; assuming that $|\chi_n(\vec b)|^2$
oscillates around this value as $n\to\infty$ we can assess the
divergence rate of the first series. Next we use the identity
\begin{equation*} 
{1\over 4\pi u}\,=\, \beta \int_0^{\infty} {e^{-\gamma u\sqrt{s}}
\over \sqrt{s}}\, ds
\end{equation*}
with $\gamma:=2\sqrt{\pi}|M|^{-1/2}$ and $\beta^{-1}:=
4\sqrt{\pi|M|}$ and write the r.h.s. as a sum of the integrals
over $(n,n\!+\!1)$ obtaining
\begin{equation*} 
\xi(\vec a;z)\,=\, \lim_{u\to 0}\, \left\lbrack
\sum_{n=0}^{\infty}\, {e^{-\kappa_n u}\over 2\kappa_n}
|\chi_n(\vec b)|^2 +\, {e^{-\gamma u\sqrt{n+1}} -e^{-\gamma
u\sqrt{n}}\over 4\pi u}\, \right\rbrack\,,
\end{equation*}
where $\kappa_n(z):= -ik_n(z)=\sqrt{\nu_n\!-\!z}$. In combination
with the preceding argument, it is easy to check that the summand
has a uniform bound of order $o(n^{-3/2})$. Hence the limit can be
interchanged with the sum and
\begin{equation} \label{xi3}
\xi(\vec a;z)\,=\, \sum_{n=0}^{\infty}\, \left\lbrack
{|\chi_n(\vec b)|^2\over 2\kappa_n(z)}  +\, {\sqrt{n}
-\sqrt{n+1}\over 2\sqrt{\pi|M|}}\, \right\rbrack\,.
\end{equation}
\begin{remarks}
(i) We do not give here details of the oscillation argument.
Notice that the conclusions made below can be obtained even
without it, up to an additive renormalization of the coupling
constant, since by the Weyl formula and the uniform boundedness of
the $\chi_n$'s the difference $\xi(\vec a;z)-\xi(\vec a;z_0)$ is
given by a convergent series.
\\ (ii) The scaling behaviour for $\Omega^{\sigma}=\mathbb{R}
\times M^{\sigma}$ with $M^{\sigma}:= \sigma M, \; \sigma>0$ is
more complicated than in the two-dimensional case. We have
$\xi(\vec{a^{\sigma}};z\sigma^{-2})= \sigma^{-1} \xi(\vec a;z)$,
so the singularities of the resolvent kernel are related by
\begin{equation} \label{scale}
\epsilon^{\sigma}(\alpha^{\sigma},\vec{a^{\sigma}}) \,=\,
\sigma^{-2} \epsilon(\alpha,\vec{a})\,, \quad \alpha^{\sigma}:=
\sigma^{-1} \alpha\,.
\end{equation}
\end{remarks}
\begin{proposition} \label{spectr1}
The operator $H(\alpha,\vec a)$ has for any $\alpha\in\mathbb{R}$
a single eigenvalue $\epsilon(\alpha,\vec a) \in(-\infty,\nu_0)$.
The corresponding eigenfunction is
\begin{equation} \label{ef1}
\psi(\vec x;\alpha,\vec a) \,=\, \sum_{n=0}^{\infty}
{e^{-\kappa_n(\epsilon)|x-a|}\over 2\kappa_n(\epsilon)}
\chi_n(\vec y) \chi_n(\vec b)\,.
\end{equation}
The function $\epsilon(\cdot,\vec a)$ is strictly increasing and
behaves as
\begin{equation} \label{weak1}
\epsilon(\alpha,\vec a) \,=\, \nu_0- \left(|\chi_0(\vec b)|^2
\over 2\alpha \right)^2 + \mathcal{O}(\alpha^{-3})\,,
\end{equation}
in the limit of weak coupling, $\alpha\to +\infty$. Moreover,
there are no eigenvalues embedded in $\sigma_c(H(\alpha,\vec a))=
[\nu_0,\infty)$.
\end{proposition}
{\sc Proof:} Due to (\ref{xi3}), $\xi(\vec a;\cdot)$ is strictly
increasing with ${\rm Ran\,}\xi= \mathbb{R}$ and $\xi(\vec a;z)=
{1\over 2} |\chi_0(\vec b)|^2 (\nu_0\!-z)^{-1/2} +\mathcal{O}(1)$
as $z\to \nu_0-$. The non-normalized eigenfunction (\ref{ef1}) is
given by the residue term in (\ref{Krein}). To check the absence
of embedded eigenvalues we have to show that $\xi(\vec a;z)=
\alpha$ has no solutions on $[\nu_0,\infty)$. Away of the
thresholds, this follows from
\begin{equation} \label{im1}
{\rm Im\,}\xi(\vec a;z)\,=\, \sum_{\{n:\, \nu_n<z\}} {|\chi_n(\vec
b)|^2 \over 2\sqrt{\nu_n\!-z}} >0\,.
\end{equation}
If $|\chi_n(\vec b)|^2\ne 0$, the resolvent kernel has a finite
limit as $z$ approaches $\nu_n$, otherwise it has the same
singularity as $G_0(\vec x_1,\vec a;\cdot)$ there, so in neither
case it has a pole. \QED \\ [2mm]
Mimicking the argument of \cite{EGST} we also get
\begin{proposition} \label{scat1}
The on-shell S-matrix at energy $z=k^2$ is a $2N_{\rm open}\times
2N_{\rm open}$ unitary matrix with elementary blocks
\begin{equation} \label{Sblock}
S_{nm}\,=\, \sqrt{k_m\over k_n} \left( \begin{array}{cc} t_{nm} &
r_{nm} \\ \tilde r_{nm} & \tilde t_{nm} \end{array} \right)\,,
\quad n,m= 1,\dots,N_{\rm open}\,,
\end{equation}
where $N_{\rm open} := {\rm card}\{ \nu_n:\: \nu_n<z\,\}$, the
tilded quantities are obtained by switching sign of the
longitudinal component of $\vec a$, $\: a\mapsto -a$, and
\begin{equation} \label{rt1}
r_{nm} e^{-ik_m a} \,=\, (t_{nm}-\delta_{nm}) e^{ik_m a} \,=\,
{i\over 2k_m}\, {e^{ik_n a}\over \alpha-\xi(\vec a;z)}\,
\chi_n(\vec b) \chi_m(\vec b)\,.
\end{equation}
\end{proposition}


\section{Finite number of perturbations}

\noindent Denote $\vec a:= \{a_1,\dots\,a_N\}$, where $\vec
a_j=(a_j,\vec b_j)$, and $\alpha:=\{ \alpha_1,\dots,\alpha_N\}$,
$j=1,\dots,N$. The Hamiltonian $(H(\alpha,\vec a)$ with $N$ point
interactions is defined as the self-adjoint extension of the
operator $-\Delta_D^M \restr C_0^{\infty}(\Omega\setminus\{\vec
a\})$ specified by the boundary conditions
\begin{equation} \label{bc N}
L_1(\psi,\vec a_j)+4\pi\alpha_j L_0(\psi,\vec a_j)\,=\,0\,, \quad
j=1,\dots,N\,.
\end{equation}
The resolvent is again found by means of the Krein formula:
\begin{equation} \label{KreinN}
(H(\alpha,\vec a)\!-\!z)^{-1}(\vec x_1,\vec x_2)\,=\, G_0(\vec
x_1,\vec x_2;z) +\, \sum_{j,k=1}^N \lambda_{jk}(\alpha,\vec a;z)
G_0(\vec x_1,\vec a;z) G_0(\vec a,\vec x_2;z)\,,
\end{equation}
where $\lambda(\alpha,\vec a;z)= \Lambda(\alpha,\vec a;z)^{-1}$
with
\begin{equation} \label{Lambda}
\Lambda_{jj}= \alpha_j-\xi(\vec a_j;z)\,, \;\quad \Lambda_{jk}=
-G_0(\vec a_j,\vec a_k;z) \quad{\rm for}\quad j\ne k\,,
\end{equation}
where $\xi(\vec a_j;z)$ is given by (\ref{xi}), (\ref{xi3}). With
these prerequisites we can derive spectral properties of our
point-interaction Hamiltonian.
\begin{theorem} \label{spectrN}
(a) The spectrum of $H(\alpha,\vec a)$ consists for any
$\alpha\in\mathbb{R}^N$ of the absolutely  continuous part $[\nu_
0,\infty)$ and eigenvalues $\epsilon_1< \epsilon_2\le \dots \le
\epsilon_m < \nu_0$ with $1\le m\le N$, given by the condition
\begin{equation} \label{poleN}
\det \Lambda(\alpha,\vec a,z)\,=\,0\,.
\end{equation}
The corresponding eigenfunctions are $ \psi(\vec x) = \sum_{j=1}^N
d_j G_0(\vec x,\vec a_j;z)$,  where $d\in\mathbb{R}^N$ solves
$\sum_{m=1}^N \Lambda(z)_{jm} d_m=0$. The ground-state
eigenfunction is positive. \\
(b) $z>\nu_0$ cannot be an eigenvalue corresponding to an
eigenvector from the subspace $\bigoplus_{\{n:\,\nu_n<z\}}
L^2(\mathbb{R})\otimes \{\chi_n\}$. On the other hand,
$H(\alpha,\vec a)$ can have embedded eigenvalues if the family
$\{\Omega,\, \vec a,\, \alpha\}$ has a suitable symmetry. \\
(c) In the weak coupling limit, $|A|:=\min_{1\le j\le N}
\alpha_j\to\infty$, there is a single eigenvalue which behaves as
\begin{equation} \label{weak ev}
\epsilon(\alpha,\vec a)\,=\, \nu_0- \left( {\left( \sum_{j=1}^N
\chi_0(\vec b_j) \right)^2 \over 2\, \sum_{j=1}^N
\alpha_j\chi_0(\vec b_j)} \right)^2 + \mathcal{O}\left(|A|^{-3}
\right)\,.
\end{equation}
\end{theorem}
{\sc Proof:} (a) A finite-rank perturbation in the resolvent
preserves $\sigma_{ac} (H_0)= [\nu_0,\infty)$. The discrete
spectrum is determined by poles of the resolvent coming from the
coefficients $\lambda_{jk}$ in (\ref{KreinN}). This yields
(\ref{poleN}); the eigenfunctions are obtained in the same way as
in \cite{AGHH,EGST}. The next question concerns the existence of
solutions to (\ref{poleN}). If $z\to -\infty$ the matrix can be
written as $\xi(\vec a;z)\tilde\Lambda(\alpha,\vec a,z)$, where
$\tilde\Lambda\to -I$, hence all eigenvalues of
$\Lambda(\alpha,\vec a,z)$ tend to $+\infty$. On the other hand,
for $\,z\to \nu_0-\,$ we have
\begin{equation*} 
\Lambda(\alpha,\vec a,z)\,=\, -\,{1\over 2\sqrt{\nu_0-z}}\,
M_1\,+\mathcal{O}(1)\,,
\end{equation*}
where $\,M_1:= (\chi_0(\vec b_j)\chi_0(\vec b_m))_{j,m=1}^N\,$.
This matrix has, in particular, an eigenvector $\,(\chi_0(\vec
b_1),\dots,\chi_0(\vec b_N))\,$ corresponding to the {\em
positive} eigenvalue $\, \sum_{j=1}^N \chi_0(\vec b_j)^2$, and
therefore at least one of the eigenvalues of $\Lambda(\alpha,\vec
a,z)\,$ tends to $\,-\infty\,$ as $\,z\to \nu_0-\,$. Using the
continuity we see that there is an eigenvalue which crosses zero,
i.e. $H(\alpha,\vec a)$ has at least one eigenvalue. By a
straightforward differentiation we find
\begin{equation*} 
{d\over dz}\, \Lambda(z)_{jm}\,=\, - \sum_{n=0}^{\infty}
{e^{-|a_j-a_m|\sqrt{\nu_n-z}} \over 4(\nu_n-z)^{3/2}}\, \left(
1+|a_j-a_m|\sqrt{\nu_n-z}\right)\, \chi_0(\vec b_j)\chi_0(\vec
b_m)\,.
\end{equation*}
The matrix function $\Lambda(\cdot)$ is monotonous if for any
$c\in\mathbb{C}^N$ the quantity ${d\over dz}(c,\Lambda(z)c)$ has a
definite sign (is non-positive in our case). This is true provided
the function $\,f:\: f(x)= e^{-\kappa|x|} (1+\kappa|x|)\,$ is of
positive type for any $\,\kappa>0\,$, which follows from the
identity
\begin{equation*} 
(1+\kappa|x|)\, e^{-\kappa|x|} \,=\, {2\kappa^3\over \pi}\,
\int_{\mathbb{R}} {e^{ipx}\over (p^2+\kappa^2)^2}\, dp
\end{equation*}
and Bochner's theorem \cite[Sec.IX.2]{RS}. In fact, since the
measure in the last integral is pointwise positive, $\,{d\over
dz}\, \Lambda(z)\,$ is even strictly positive; it means that all
the eigenvalues of $\,\Lambda(\alpha,\vec a;z)\,$ are decreasing
functions of $\,z\,$ and $H(\alpha,\vec a)$ has at most $N$
eigenvalues.

To check that the ground state is non-degenerate, we have to
demonstrate that the lowest eigenvalue of $\Lambda(z)$ is simple
for any $z\in (-\infty,\nu_0)$, which is equivalent to the claim
that the matrix semigroup $ \lbrace\, e^{-t\Lambda(z)} :\, t\ge
0\,\rbrace$ is positivity preserving \cite[Sec.XIII.12]{RS}. The
last property is ensured if all the non-diagonal elements of
$\Lambda(z)$ are negative; we have $\Lambda(z)_{jm}= -G_0(\vec
a_j,\vec a_m;z)$ by (\ref{Lambda}) so the desired result follows
from the positivity of the free-resolvent kernel. The coefficients
may be therefore chosen of the same sign for the ground state; in
fact, as strictly positive because $d_{j_0}=0$ would mean that the
eigenfunction is smooth at $\vec x=\vec a_{j_0}$ so the
corresponding interaction is absent, $\alpha_{j_0}= \infty$.

(b) Suppose now that $H\varphi= z\varphi$ for some $z>\nu_0$. We
adapt again the argument from \cite[Sec.II.1]{AGHH} and pick an
arbitrary $z'\in\rho(H)$; then there is a vector $\psi_0\in
D(H_0)$ which allows us to write
\begin{equation} \label{eigenvector decomposition}
\varphi\,=\, \psi_0+\, \sum_{j=1}^N d_j G_0(\cdot,\vec a_j;z')\,.
\end{equation}
Furthermore, we expand $\psi_0$ as a series, $\psi_0(\vec x)=
\sum_{n=0}^{\infty} g_n(x) \chi_n(\vec y)\,$ with the coefficients
$g_n\in L^2(\mathbb{R})$. Using the identity $(H_0-z)\psi_0=
(z-z') \sum_{j=1}^N d_j G_0(\cdot,\vec a_j;z')$ and the fact that
$\{ \chi_n\}$ is an orthonormal basis in $L^2(M)$, we obtain a
system of equations; by the Fourier-Plancherel operator it is
transformed into
\begin{equation} \label{g hat}
(p^2-z+\nu_n)\hat g_n(p)\,=\, {z-z'\over 2\pi}\, \sum_{j=1}^N d_j
\chi_n(\vec b_j)\, {e^{-ipa_j}\over p^2-z'+\nu_n} \,.
\end{equation}
If $g_n\in L^2$ the same has to be true for $\hat g_n$; this is
impossible if $z>\nu_n$ and the r.h.s. of (\ref{g hat}) is nonzero
at $\pm p_n$, where $p_n:= \sqrt{\nu_n\!-z}$, since $\hat g_n^2$
would have then a non-integrable singularity. If $N>1$, it might
happen that the r.h.s. of (\ref{g hat}) is not zero identically.
However, if the $a_j$ are mutually different, $\sum_{j=1}^N
d_j\chi_n(\vec b_j)\, e^{\mp ipa_j}=0$ implies $d_j=0$ by linear
independence. On the other hand, if some of them coincide we find
$\sum_j d_j\chi_n(\vec b_j)=0$ where the index runs through the
values with the same longitudinal coordinate $a_j$, and therefore
$\hat g_n=0$ again.

The condition $\nu_n<z$ in the above argument is crucial; the
operator $\,H(\alpha,\vec a)\,$ can have embedded eigenvalues with
eigenfunctions in the orthogonal complement of the mentioned
subspace if $N>1$. Examples can be constructed as in \cite{EGST}
(or other similar systems -- cf.\cite{ELV}) using $M$ with a
symmetry: one has to choose a family of weak enough point
interactions with the same symmetry.

(c) We denote $\,A:= {\rm diag}(\alpha_1,\dots,\alpha_N)$, and use
the decomposition
\begin{equation*} 
\Lambda(z)\,=\, \left( A-\tilde\Gamma(z) \right) \left\lbrack\, I-
\left( A-\tilde\Gamma(z) \right)^{-1} {M_1\over 2\sqrt{\nu_0-z}}\,
\right\rbrack\,,
\end{equation*}
where $\tilde\Gamma(z)$ is a remainder independent of
$\,\alpha\,$, whose norm is bounded as $z\to \nu_0-$, and $M_1$ is
the matrix defined above. The first factor is regular for $|A|$
large enough. Since $M_1$ is rank one we have to solve the
equation
\begin{equation*} 
\eta \sum_{j,k=1}^N \chi_0(\vec b_j) \alpha_j \left(I-
\tilde\Gamma(z)A^{-1} \right)_{jk} \chi_0(\vec b_k) - \sum_{j=1}^N
\chi_0(\vec b_j)^2\,=\,0
\end{equation*}
with $\eta:= 2\sqrt{\nu_0\!-z}$; then (\ref{weak ev}) follows by
the implicit-function theorem. \QED
\begin{remarks}
(i) We get also the weak-coupling asymptotics for the
eigenfunction:
\begin{eqnarray*}
\psi(x;\alpha,\vec a) &\!\approx\!& \chi_0(\vec y) \:
{2\sum_{j=1}^N \alpha_j\chi_0(\vec b_j)^2 \over \left(
\sum_{j=1}^N \chi_0(\vec b_j)^2 \right)^2}\: \sum_{j=1}^N
e^{-\sqrt{\nu_0-\epsilon}|x-a_j|} \chi_0(\vec b_j)^2
\\ \\ && +\, \sum_{n=1}^{\infty}\, \chi_n(\vec b_j)\: \sum_{j=1}^N
{e^{-\sqrt{\nu_n-\nu_0}|x-a_j|} \over \sqrt{\nu_n-\nu_0}}\,
\chi_n(\vec b_j) \chi_0(\vec b_j)\,.
\end{eqnarray*}
The leading term is a product of $\,\chi_1(\vec y)\,$ with a
linear combination of the eigenfunctions of one-dimensional point
interactions placed at $a_j,\; j=1,\dots,N\,$.
\\ (ii) The scattering problem can be treated as in the one-center
case. Existence and completeness of wave operators follow from the
Kato--Birman theory \cite{RS}. The reflection and transmission
amplitudes from the $n$-th to the $m$-th channel are
\begin{eqnarray} \label{rt_N}
r_{nm}(z) &\!=\!& {i\over 2}\, \sum_{j,k=1}^N
(\Lambda(z)^{-1})_{jk} {\chi_m(\vec b_j) \chi_n(\vec b_k)\over
k_m(z)}\, e^{i(k_ma_j+k_na_k)}\;, \nonumber \\ t_{nm}(z) &\!=\!&
\delta_{nm}\,+\, {i\over 2}\, \sum_{j,k=1}^N
(\Lambda(z)^{-1})_{jk} {\chi_m(\vec b_j) \chi_n(\vec b_k)\over
k_m(z)}\, e^{-i(k_ma_j-k_na_k)}
\end{eqnarray}
and the unitarity condition now reads
\begin{eqnarray} \label{unitarity_N}
\sum_{\{m:\: 0\le \nu_m< z\}} k_m (t_{nm}\overline
t_{sm}+r_{nm}\overline r_{sm}) &\!=\!& \delta_{ns} k_n\,,
\nonumber \\ \sum_{\{m:\: 0\le \nu_m< z\}} k_m \left(\tilde
t_{nm}\overline r_{sm} +\tilde r_{nm}\overline t_{sm}\right)
&\!=\!& 0 \,,
\end{eqnarray}
because $\,S_{nm}\,$ is given again by (\ref{Sblock}), where the
tilded quantities are obtained by mirror transformation, $\,a_j\to
-a_j\,$.
\end{remarks}


\section{The periodic case}

\noindent  In the infinite-center case we restrict ourselves to
the periodic situation, i.e. we suppose that the set
$\{\alpha,\vec a\}_{per}=\{[\alpha_j,\vec a_j]:\; j=1,2,\dots\,\}$
in $\Omega$ is countably infinite and has a periodic pattern with
a period $\ell>0$ and $N$ perturbations in each cell, which we
denote again as $\{\alpha,\vec a\}$. Following the Floquet-Bloch
decomposition, we find the unitary operator $U:\, L^2(\Omega)\to
L^2(\mathcal{B}, (\ell/2\pi)d\theta; L^2(\hat\Omega))$, where
\begin{equation} \label{WS cell}
\hat\Omega\,:=\, [0,\ell)\times M\,, \qquad \mathcal{B}\,:=\,
\left\lbrack\, -{\pi\over\ell},\,{\pi\over\ell}\right) \times M
\;;
\end{equation}
the $x$-projections of these sets are the Wigner-Seitz cell of the
underlying one-dimensional lattice and the corresponding Brillouin
zone, respectively. By means of $U$, the operator $H(\alpha,\vec
a)$ is unitarily equivalent to
\begin{equation} \label{FB decomposition}
U\,H(\{\alpha,\vec a\}_{per})\,U^{-1}\,=\, {\ell\over 2\pi}\,
\int_{|\theta\ell|\le\pi}^{\oplus} H(\alpha,\vec
a;\theta)\,d\theta\,,
\end{equation}
where $H(\alpha,\vec a;\theta)$ is the point-interaction
Hamiltonian on $L^2(\hat\Omega)$, i.e. the Laplacian satisfying
(\ref{bc N}) at the points $\vec a_j$, Dirichlet b.c. for $x\in
[0,\ell)\,, \: \vec y\in\partial M$, and
\begin{equation} \label{Bloch bc}
\psi(\ell-,\vec y)\,=\,e^{i\theta\ell} \psi(0+,\vec y)\,, \qquad
{\partial\psi\over\partial x}(\ell-,\vec y)\,=\,e^{i\theta\ell}\;
{\partial\psi\over\partial x}(0+,\vec y)
\end{equation}
for $\vec y\in M$. The resolvent of $H(\alpha,\vec a;\theta)$ can
be derived by modifying the argument of \cite[Sec.5]{EGST}. The
``free'' eigenvalues
\begin{equation} \label{FB free eigenvalues}
\epsilon_{mn}(\theta)\,:=\, \left({2\pi
m\over\ell}\,+\theta\right)^2 +\nu_n\,, \quad m\in\mathcal{Z}\,,\;
n=0,2,\dots\,,
\end{equation}
correspond to the eigenfunctions $\eta^{\theta}_m\otimes\chi_n$,
where $\chi_n$ are as above and $\eta^{\theta}_m(x):=
\ell^{-1/2}\, e^{i(2\pi m+\theta\ell)x/\ell },\, m\in\mathcal{Z}$.
Moreover, the free resolvent kernel is in analogy with \cite{EGST}
obtained by a partial summation of the appropriate double series
and equals
\begin{eqnarray} \label{FB free resolvent}
G_0(\vec x_1,\vec x_2;\theta;z) &\!=\!& \sum_{n=0}^{\infty}\,
{\sinh((\ell\!-\!|x_1\!-\!x_2|) \sqrt{\nu_n\!-\!z})
+e^{2i\eta\theta\ell} \sinh(|x_1\!-\!x_2| \sqrt{\nu_n\!-\!z})
\over \cosh(\ell\sqrt{\nu_n\!-\!z}) -\cos(\theta\ell)} \nonumber
\\ && \times\:{\chi_n(\vec y_1) \chi_n(\vec y_2) \over
2\sqrt{\nu_n\!-\!z}}\,,
\end{eqnarray}
where $\eta:={\rm sgn}(x_1\!-x_2)$. The full kernel is then
expressed by a formula analogous to (\ref{KreinN}) with
\begin{equation} \label{FB lambda}
\lambda(\alpha,\vec a,\theta;z)\,=\, \Lambda(\alpha,\vec
a,\theta;z)^{-1}\,,
\end{equation}
where $\Lambda_{jr}= -G_0(\vec a_j,\vec a_r)$ for $j\ne r$, while
the diagonal elements are given by
\begin{equation} \label{FB Lambda}
\Lambda_{jj} \,=\, \alpha_j\,-\,{1\over 2}\, \sum_{n=0}^{\infty}\,
\left( {\sinh(\ell\sqrt{\nu_n\!-\!z}) \over
\cosh(\ell\sqrt{\nu_n\!-\!z}) -\cos\theta\ell}\: {\chi_n(\vec
y)^2\over \sqrt{\nu_n\!-\!z}}\,+\, {\sqrt{n}- \sqrt{n\!+\!1}\over
\sqrt{\pi|M|}} \right)\,.
\end{equation}
We may also write the last formula as $\Lambda_{jj}(\alpha,\vec
a,\theta;z) =\alpha_j- \xi(\vec a_j,\theta;z)$, where the function
$\xi$ is for $z\in\mathbb{R}$ more explicitly given by
\begin{eqnarray} \label{FB xi}
\xi(\vec a_j,\theta;z) &\!=\!& {1\over 2}\, \sum_{\{n:\: \nu_n\le
z\}}\, \left( {\sin(\ell\sqrt{z\!-\!\nu_n}) \over
\cos(\ell\sqrt{z\!-\!\nu_n}) -\cos\theta\ell}\: {\chi_n(\vec
b_j)^2\over \sqrt{z\!-\!\nu_n}}\,+\, {\sqrt{n}-
\sqrt{n\!+\!1}\over \sqrt{\pi|M|}} \right) \nonumber \\ &\!+\!&
{1\over 2}\, \sum_{\{n:\: \nu_n>z\}}\, \left(
{\sinh(\ell\sqrt{\nu_n\!-\!z}) \over \cosh(\ell\sqrt{\nu_n\!-\!z})
-\cos\theta\ell}\: {\chi_n(\vec b_j)^2\over
\sqrt{\nu_n\!-\!z}}\,+\, {\sqrt{n}- \sqrt{n\!+\!1}\over
\sqrt{\pi|M|}} \right)\,.
\end{eqnarray}
It is defined everywhere except at
\begin{equation} \label{FB singularities}
\mathcal{E}(\vec a,\theta)\,:=\, \{\,
\epsilon_{mn}(\theta)\in\mathcal{E}(\theta)\,:\; \chi_n(\vec
b_j)\ne 0\;\}\,,
   \end{equation}
where $\mathcal{E}(\theta)$ is the eigenvalue set (\ref{FB free
eigenvalues}). Of course, the r.h.s. of (\ref{FB singularities})
makes no sense if $z=\nu_n$ and $\chi_n(\vec b_j)\ne 0$, but
applying general results on self-adjoint extensions to the
one-center case, we can establish {\em a posteriori} that $\xi$
can be defined there by continuity. Moreover, we find that the
function $\xi(\vec a_j,\theta;\cdot)$ is monotonously increasing
between any pair of neighboring singularities. Then we have the
following result.
\begin{proposition} \label{spec-per}
For a given $\alpha\in\mathbb{R}^N$ the operator $H(\alpha,\vec
a;\theta)$ has $N$ eigenvalues $\epsilon_{mnj}(\alpha,\vec
a;\theta),\: j=1,\dots,N$, in any gap of the set (\ref{FB
singularities}) determined by
\begin{equation} \label{FB pole}
\det \Lambda(\alpha,\vec a,\theta;z)\,=\,0\,.
\end{equation}
The corresponding eigenfunctions are $\psi(\vec x) = \sum_{j=1}^N
d_j G_0(\vec x,\vec a_j,\theta;z)$ where the $d_j$'s are
determined by $\Lambda(\alpha,\vec a,\theta;z)$ as in
Theorem~\ref{spectrN}.
\end{proposition}
The spectrum of the original Hamiltonian $H(\{\alpha,\vec
a\}_{per})$ consist then of bands,
\begin{equation} \label{spec-band}
\sigma\left(H(\{\alpha,\vec a\}_{per})\right)\,=\, \bigcup_{mnj}
\left\{\, \epsilon_{mnj}(\alpha,\vec a;\theta):\: \theta\in
\mathcal{B}\, \right\}\,.
\end{equation}
One is interested, of course, in its absolute continuity and
existence of gaps. We restrict ourselves to the simplest
nontrivial situation.
\begin{example}
Let $N=1$, i.e. let each cell contain a single point interaction.
The condition (\ref{FB pole}) then simplifies to
\begin{equation} \label{FB pole_1}
\xi(\vec a,\theta;z)\,=\,\alpha\,.
\end{equation}
The left hand side is monotonously increasing between its
singularities, i.e. the points of $\mathcal{E}(\vec a,\theta)$.
This means that for fixed $\alpha,\,\theta$ there is a sequence
$\{\epsilon_r(\alpha,\vec a,\theta)\}_{r=0} ^{\infty}$ arranged in
the ascending order; each of them depends, in fact, only on the
transverse component $\vec b$ of the vector $\vec a$. The lowest
one satisfies
\begin{equation*} 
\epsilon_0(\alpha,\vec a,\theta)\,<\,\nu_0+\theta^2
\end{equation*}
and between each two neighboring points of $\mathcal{E}(\vec
a,\theta)$ there is just one of the other eigenvalues. It is also
clear that any of $\epsilon_r(\alpha,\vec a,\theta)$ is continuous
with respect to the parameters and $\epsilon_r(\cdot,\vec
a,\theta)$ is increasing for fixed $\vec b$ and $\theta$.
Concerning the $\theta$-dependence, the implicit-function theorem
tells us that
\begin{equation*} 
{\partial\epsilon_r(\alpha,\vec a,\theta)\over
\partial\theta}\,=\, -\, \left.{\partial\xi(\vec a,\theta;z)\over
\partial\theta}\, \left( {\partial\xi(\vec a,\theta;z)\over
\partial z} \right)^{-1}\, \right|_{(\epsilon_r,\theta)}
\end{equation*}
whenever the denominator is nonzero. Away of the thresholds,
$z=\nu_n$, and the points of $\mathcal{E}(\theta)$, a
straightforward differentiation shows that $\xi(\vec
a,\cdot;\cdot)$ is analytic in both variables. By (\ref{FB xi})
the numerator is not identically zero; hence the derivative
$\partial\epsilon_r(\alpha,\vec a,\theta)/
\partial\theta$ may be zero at some points but never in an
interval and {\em the spectrum of $H(\{\alpha,\vec a\}_{per})$ is
absolutely continuous} \cite[Sec.XIII.16]{RS}.

Let us turn now to the question about the number of gaps. Below
$z=\nu_0$ the spectrum may be estimated by means of extrema of the
function $\xi$ which yield $\theta$-independent bounds: we have
$\xi(\vec a,\theta;z) \le \xi_+(\vec a,z)$ where
\begin{equation*} 
\xi_+(\vec a,z):= \max_{|\theta\ell|\le\pi} \xi(\vec a,\theta;z)
\,=\, {1\over 2} \sum_{n=0}^{\infty}\,\left( {\chi_0(\vec b)^2
\over \sqrt{\nu_n\!-\!z}}\, \coth\left({\ell\over
2}\sqrt{\nu_n\!-\!z} \right) \,+\, {\sqrt{n}-\sqrt{n\!+\!1}\over
\sqrt{\pi|M|}} \right)
\end{equation*}
and a similar formula for the minimum, $\xi_-(\vec a,z)$, with
$\,\coth\,$ replaced by $\,\tanh$. Both functions are continuously
increasing and tend to $-\infty$ as $z\to -\infty$. On the other
hand, $\xi_+(\vec a,\cdot)$ diverges as $z\to \nu_0-$ while the
lower bound $\xi_-(\vec a,\cdot)$ has a finite limit. This shows,
in particular, that the spectral condition (\ref{FB pole_1}) has
no solution for any $\theta$ in a left neighborhood of $\nu_0$
provided
\begin{eqnarray} \label{one gap}
\alpha\!&<&\! \xi_-(\vec a,\nu_0-) \,=\, {\ell\over 4}\,
\chi_0(\vec b)^2 -\,{1\over 2\sqrt{\pi|M|}} \nonumber\\ &\!+\!&
{1\over 2}\,\sum_{n=1}^{\infty}\,\left( {\chi_0(\vec b)^2 \over
\sqrt{\nu_n\!-\!\nu_0}}\, \tanh\left({\ell\over
2}\sqrt{\nu_n\!-\!\nu_0} \right) \,+\,
{\sqrt{n}-\sqrt{n\!+\!1}\over \sqrt{\pi|M|}} \right)\,;
\end{eqnarray}
in other words, that a gap exists. The condition (\ref{one gap})
is satisfied for a strong enough coupling if the point-interaction
spacing is kept fixed. On the other hand, inspecting the
right-hand-side we see that the gap exists also for any fixed
$\alpha$ and the spacing $\ell$ large enough. In this respect the
spectrum is similar to that of a straight polymer in
$\mathbb{R}^3$ described in \cite[Sec.III.1]{AGHH}. However, for
our ``coated polymer'' a much stronger result is valid: we shall
show that under a suitable choice of parameters it can have {\em
any finite number} of gaps.

To this end, we consider $z\in(\nu_0+\varepsilon,
\nu_1-\varepsilon)$ for a fixed $\varepsilon>0$ and $\ell\gg
\sqrt{|M|}$, and rewrite the right hand side of the relation
(\ref{FB xi}) as
\begin{equation*} 
\xi(\vec a,\theta;z)\,=\, \xi_0(\vec a,\theta;z)+ \eta(\vec
a,\theta;z)\,,
\end{equation*}
where
\begin{equation*} 
\xi_0(\vec a,\theta;z)\,:=\, {\sin(\ell\sqrt{z\!-\!\nu_0}) \over
\cos(\ell\sqrt{z\!-\!\nu_0}) -\cos\theta\ell}\; {\chi_0(\vec
b)^2\over 2\sqrt{z\!-\!\nu_0}}
\end{equation*}
and $\eta(\vec a,\theta;z)$ is the rest. The latter is
monotonously increasing with respect to z and its derivative is
bounded everywhere below the second threshold, in particular, in
the chosen interval of energies. Moreover, $\eta(\vec a,\theta;z)$
is bounded from above by
\begin{equation*} 
\eta_+(\vec a,z)\,:=\, -\,{1\over
2\sqrt{\pi|M|}} \,+\, {1\over 2} \sum_{n=1}^{\infty}\,\left(
{\chi_0(\vec b)^2 \over \sqrt{\nu_n\!-\!z}}\,
\coth\left({\ell\over 2}\sqrt{\nu_n\!-\!z} \right) \,+\,
{\sqrt{n}-\sqrt{n\!+\!1}\over \sqrt{\pi|M|}} \right)
\end{equation*}
and the corresponding minimum, $\eta_-(\vec a,z)$, is obtained
when $\,\coth\,$ in the last expression is replaced by $\,\tanh$.
These estimates shrink as $\ell$ becomes large: using the
inequality $\coth u-\tanh u< 5\, e^{-2u}$ for $2u\ge 1$, we find
\begin{equation*} 
\eta_+(\vec a,z)-\eta_-(\vec a,z)\,<\, {5\over 2}\, {\chi_1(\vec
b)^2 \over \sqrt{\nu_1\!-\!z}}\, e^{-\ell\sqrt{\nu_1\!-\!z}} \,+\,
{5\over 2}\, \sum_{n=2}^{\infty}\, {\chi_n(\vec b)^2 \over
\sqrt{\nu_n\!-\!z}}\, e^{-\ell\sqrt{\nu_n\!-\!z}}\,.
\end{equation*}
The series can be estimated by an integral, which yields for a
fixed $z\in(\nu_0+\varepsilon, \nu_1-\varepsilon)$ the behavior
\begin{equation} \label{corridor2}
\eta_+(\vec a,z)-\eta_-(\vec a,z)\,=\,
\mathcal{O}\left(\sqrt{|M|}\, \ell^{-1} \right)\,.
\end{equation}
On the other hand, the function $g_{\theta}(u):= \sin u (\cos
u-\cos\theta\ell)^{-1}$ is increasing between any two zeros of its
denominator. In the intervals, where it is positive, it can be
estimated from below by the appropriate branch of $\tan\left(
{u\over 2}+\pi m\right)$; when it is negative, we have a similar
estimate from above with $\,\tan\,$ replaced by $\,-\cot$. Hence
independently of $\theta$ we have either
\begin{equation*} 
\xi_0(\vec a,\theta;z)\,\ge\, {\chi_n(\vec b)^2\over
2\sqrt{z\!-\!\nu_0}}\; \tan\left( {\pi\over 2}\,\left\lbrace
{\ell\over\pi} \sqrt{z\!-\!\nu_0} \right\rbrace \right)
\end{equation*}
or
\begin{equation*} 
\xi_0(\vec a,\theta;z)\,\le\, -\,{\chi_n(\vec b)^2\over
2\sqrt{z\!-\!\nu_0}}\; \cot\left( {\pi\over 2}\,\left\lbrace
{\ell\over\pi} \sqrt{z\!-\!\nu_0} \right\rbrace \right)\,,
\end{equation*}
where $\{\cdot\}$ denotes the fractional part. Putting the
estimates together we see that the oscillating part dominates, so
for sufficiently large $|\alpha|$ there are gaps having
$\nu_0+\,\left( \pi m\over\ell\right)^2$ as one endpoint provided
it belongs to $(\nu_0+\varepsilon, \nu_1-\varepsilon)$. In
addition, $\tan u+\cot u\ge 2$ which means that the gap between
the lower and the upper bound to $\xi(\vec a,\theta;z)$ never
closes within the chosen interval if $\ell/\sqrt{|M|}$ is large
enough; we infer that {\em for any} $\alpha\in\mathbb{R}$ the
operator $H(\{\alpha,\vec a\}_{per})$ can have {\em an arbitrary
finite number of gaps} in its spectrum provided the spacing of the
point-interaction array is large enough.
\end{example}
Conclusions of the example can be extended to a finite number of
point perturbations per cell. In a similar way one can treat a
toroidal tube supporting point interactions and threaded by a
magnetic flux, etc.

A more difficult question concerns the finiteness of the number of
open gaps. Recall that the gap number depends strongly on the
dimension of a periodic system: for one-dimensional systems it is
generically infinite \cite[Sec.XIII.16]{RS},
\cite[Sec.III.2]{AGHH}, while for higher dimensions it is finite
by the {\em Bethe-Sommerfeld conjecture}. The latter is known to
be true, in particular, for periodic potentials or lattices of
point interactions in $\mathbb{R}^3$ -- see \cite{AG,Sk}. The tube
boundary makes things more complicated, but one still expects that
at high energies gaps will close due to overlapping of
contributions from different transverse modes. Notice in this
connection that in the example we have been looking for gaps in
the energy interval where transport is possible in the lowest
transverse mode only. Nevertheless, it is not easy to demonstrate
that no gaps remain open above a certain energy.

In a similar way, the ``mixed dimensionality" of waveguide systems
inspires other questions such as existence of a mobility edge in
tubes with random point interactions, etc. Generally speaking,
solvable models whose genealogy can be traced back to the treatise
\cite{AGHH} will represent for long a useful laboratory for the
spectral theory.

\end{document}